# Combined approach for automatic and robust calculation of dominant frequency of electrogastrogram

Nebojša Jovanović[a,*], Nenad B. Popović[a], Nadica Miljković[a]

[a] University of Belgrade - School of Electrical Engineering, Bulevar kralja Aleksandra 73, 11000 Belgrade, Serbia

E-mail addresses:

Nebojša Jovanović, nebojsa.php@gmail.com

Nenad B. Popović, nenad.pop92@gmail.com

Nadica Miljković, nadica.miljkovic@etf.rs

*Corresponding Author:

Nebojša Jovanović, student

University of Belgrade - School of Electrical Engineering

Bulevar kralja Aleksandra 73, 11000 Belgrade, Serbia

e-mail: nebojsa.php@gmail.com

tel.: +381 11 3218-348

## Abstract

We present a novel method for automatic and robust detection of dominant frequency (DF) in the electrogastrogram (EGG). Our new approach combines Fast Fourier Transform (FFT), Welch's method for spectral density estimation, and autocorrelation. The proposed combined method as well as other separate procedures were tested on a freely available dataset consisted of EGG recordings in 20 healthy individuals. DF was calculated in relation (1) to the fasting and postprandial states, (2) to the three recording locations, and (3) to the subjects' body mass index.

For the estimation of algorithms performance in the presence of noise, we created a synthetic dataset by adding white Gaussian noise to the artifact-free EGG waveform in one subject. The individual algorithms and novel combined approach were evaluated in relation to the signal-to-noise ratio (SNR) in range from -40 dB to 20 dB.

Our results showed that the novel combined method significantly outperformed the commonly used approach for DF calculation – FFT in noise presence when compared to the benchmark data being was manually corrected by an expert. The novel method outperformed autocorrelation and Welch's method in accuracy. Additionally, we presented a method for optimal window width selection when using Welch's spectrogram that showed that for DF detection, window length of *N/4* (300 s), where *N* is the length of EGG waveform in samples, performed the best when compared to the benchmark data.

The combined approach proved efficient for automatic and robust calculation of dominant frequency on openly available EGG dataset recorded in healthy individuals and is promising approach for DF detection.

**Keywords**: automatic analysis, BMI, dominant frequency, electrogastrography, FFT

# 1 Introduction

Surface electrogastrography (EGG) applies cutaneous electrodes placed over muscle belly for electrical gastric activity assessment [1]. The main component of EGG signal is the slow-wave at a frequency of approximately 3 cycles per minute (cpm) or 0.05 Hz in healthy subjects [2]. Analysis of EGG signals recording in fasting and postprandial states is commonly used to evaluate gastric motility [2]. Also, EGG recordings can be used to diagnose gastric disorders such as nausea, vomiting, functional dyspepsia, gastroparesis, motion sickness, etc. [3-7]. Due to its noninvasive nature EGG could be used for a wide range of applications in both research and clinical conditions. One of the main obstacles in wider application of surface EGG is its exposure to the noise, such as respiratory artifacts, electrocardiography (ECG) interference, motion artifacts, etc. [8]. Electrical activities from other parts of the gastrointestinal (GI) system, such as the colon, can also overlap with normal rhythmical gastric activity and deteriorate signal quality [9].

The most commonly calculated EGG feature is the dominant frequency (DF) and it is considered to be the main parameter of the EGG, alongside dominant power (DP) [10]. DF defines the highest frequency of stomach contractions and can indicate the presence of abnormal gastric rhythms i.e., bradygastria (DF in 1-2 cpm range) or tachygastria (DF in 4-10 cpm range) [11]. Commonly, DF is calculated as the global maximum in the frequency spectrum of the filtered EGG signal [12]. Owing to the vulnerability of EGG signal to artifacts, researchers commonly apply visual inspection of the recorded EGG data in order to perform an appropriate manual correction for large artifacts and to verify the quality of the filtering process in order to be sure that DF was calculated appropriately [13].

In this paper, we evaluate three procedures for DF calculation and propose an automatic combined method for calculation. Our idea is based on the fact that combined approach would yield the best performance. Therefore, we combined Welch's spectrum and autocorrelation as we suggested that they would increase efficiency in a noisy environment, and FFT in order keep spectral resolution at the satisfactory level.

## 1.2 Study aim

We tested four algorithms (autocorrelation, Welch's spectrum, FFT, and novel combined auto-method abbreviated as NCAM) on open and semi-synthetic EGG data, consisting of EGG recordings performed in 20 healthy volunteers. The analysis was done in relation to the recording states (fasting and postprandial), to the three recording electrode locations, and body mass index (BMI) with two groups of BMI (low and high). Also, we investigated the novel method for DF calculation in relation to the signal-to-noise ratio (SNR) in order to assess its robustness to noise.

The main aim of the study was to examine three selected methods for the automatic calculation of DF and to propose and evaluate a novel combined method NCAM for automatic DF calculation.

## 2 Methods and materials

For evaluation of the proposed method, we used freely available data, but for assessment of algorithm performance in the presence of artifacts, we generated a semi-synthetic dataset.

### 2.1 EGG data

The dataset consists of EGG recordings in 20 healthy volunteers (8 Females and 12 Males) [14, 15]. For each subject, three-channel EGG was recorded (from CH1, CH2, and CH3 locations), during fasting and postprandial states (20 min duration each). The subjects were divided in relation to BMI into two groups - low BMI subjects (BMI < 25 kg/m²) and high BMI subjects (BMI > 25 kg/m²). More details regarding the study group and the recording protocol can be found in [14].

All processing steps were performed in Python programming language (Python Software Foundation, Delaware, USA). The following Python packages were used: NumPy, SciPy, pandas, matplotlib, and seaborn [16-23].

Firstly, signals were filtered with Butterworth third-order bandpass filter with cutoff frequencies of 0.03 Hz and 0.25 Hz in order to remove artifacts and to preserve slow wave component in EGG signal. Digital filtering was performed with the function scipy.signal.filtfilt in order to avoid phase distortion [18]. Then, the DF was calculated.

### 2.2 Dominant frequency calculation

Normalized direct autocorrelation $\rho_k$ of digital EGG signal with samples $X_i$ where index $i$ is in the range from 0 to $N$-1 ($N$ is the signal length of the discrete signal) at lag $k$ was computed by the following formula [24]:

$$\rho_k = \frac{\sum_{i=k}^{N-1} X_i X_{i-k}}{\sqrt{\sum_{j=k}^{N-1} X_j^2} \sqrt{\sum_{j=0}^{N-k-1} X_j^2}} \quad (1)$$

Normalized autocorrelations for lags from $k_1$ = 25 samples to $k_2$ = 67 samples were calculated for all EGG signals. Lags $k_1$ and $k_2$ correspond to frequencies of 4.8 cpm and 1.8 cpm, respectively, being in the range which contains the normal gastric rhythm. The first peak at a value greater than zero was determined and the $DF_{AC}$ was calculated as the ratio presented in Relation (2), where $index_{max}$ is the lag at the first peak greater than zero and $f_s$ is the sampling frequency (2 Hz).

$$DF_{AC} = \frac{fs}{index_{max}} \quad (2)$$

For estimating DF from the Power Spectral Density (PSD), we used function proposed by Welch [25]. The method involves sectioning the data record and averaging modified periodograms of the sections. For implementation of Welch's method we used existing scipy.signal.welch function in scipy Python package [18]. The length of the window was selected empirically for each of the three channels with ratio *S* ranging from *S* = 1.5 to *S* = 11 with step of 0.5. *S* is the ratio of the signal and window lengths. For all *S* values, a paired-sample t-test was performed and the p-value was determined for each of the three channels for fasting and postprandial states. The one at which the average p-value across all three channels is closest to the p-values obtained for presenting the difference between fasting and postprandial in Popović et al. [14] was selected to be the optimal window length. For *noverlap* parameter, half of the window length was chosen, and we set the number of points for PSD calculation at 2048. In the end, the frequency of the peak value from the chosen window length (global maximum) was determined by the following Relation (3):

$$DF_{Welch} = \arg\max_f PSD(f) \tag{3}$$

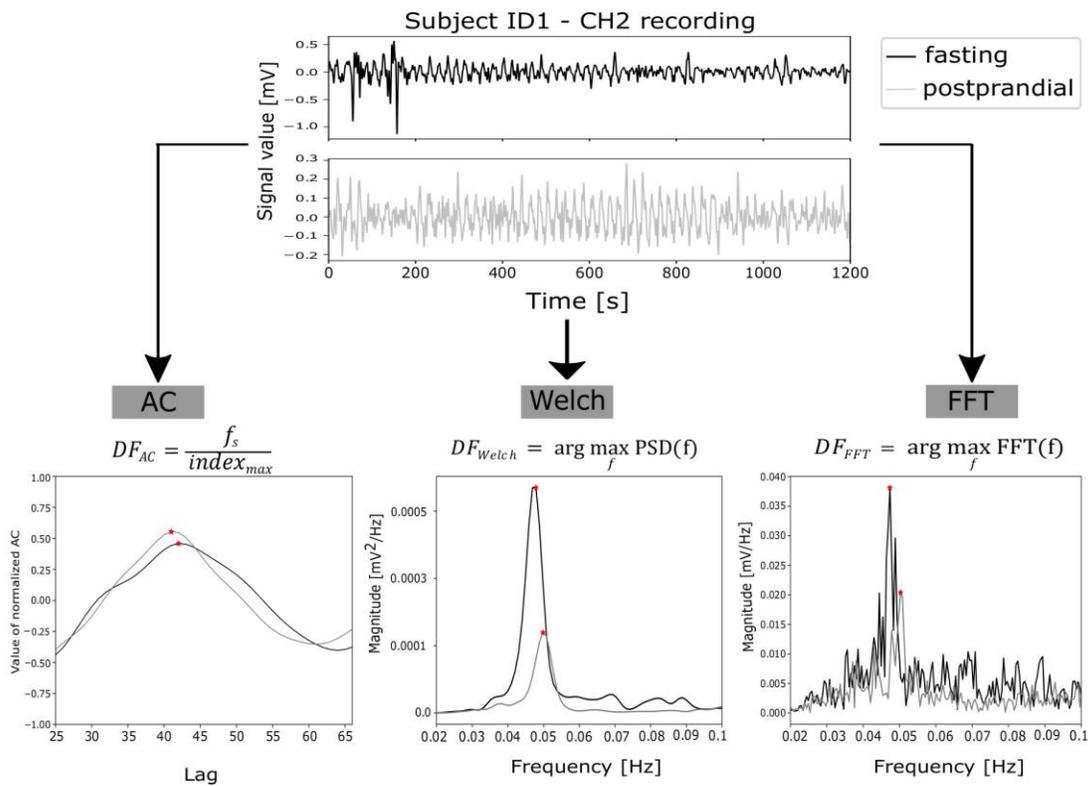

Figure 1. EGG signal in subject ID1 from the database [15] during fasting and postprandial states with corresponding autocorrelations, FFTs, and Welch's spectrograms and for signal recorded from CH2 location. Formulas for calculation of dominant frequencies are shown above the graphs where $DF_{AC}$, $DF_{Welch}$, and $DF_{FFT}$ represent the dominant frequencies calculated by autocorrelation, Welch's method, and FFT, respectively. $f_s$ is sampling frequency and $index_{max}$ is the lag at which autocorrelation function has its first peak. The fasting state is presented in black and the postprandial state is presented in light grey.

FFT was computed and the frequency at the peak value (global maximum) was determined as in Relation (4). The number of points for FFT calculation was set at 4096. We did not provide manual correction of the DF calculated by FFT.

$$DF_{FFT} = \arg\max_{f} FFT(f) \qquad (4)$$

Calculation of DF by autocorrelation, FFT, and Welch's method by Relations (2)-(4), for sample EGG data during postprandial and fasting states is presented in Fig. 1.

## 2.2.1 Proposed novel method (NCAM)

The proposed method is presented in Fig. 2. If the normalized autocorrelation peak is higher than 0.4, we hypothesize that the spectral peak in FFT is the true DF as it is prominent due to its power that corresponds to relatively higher SNR. If the peak is lower than 0.4, we suggested a new formula for calculating DF given in Relation (5), which uses all previously described methods.

$$DF = DF_{Welch} + peak_{AC} \times \left(DF_{FFT} - \frac{f_s}{index_{max}}\right) \qquad (5)$$

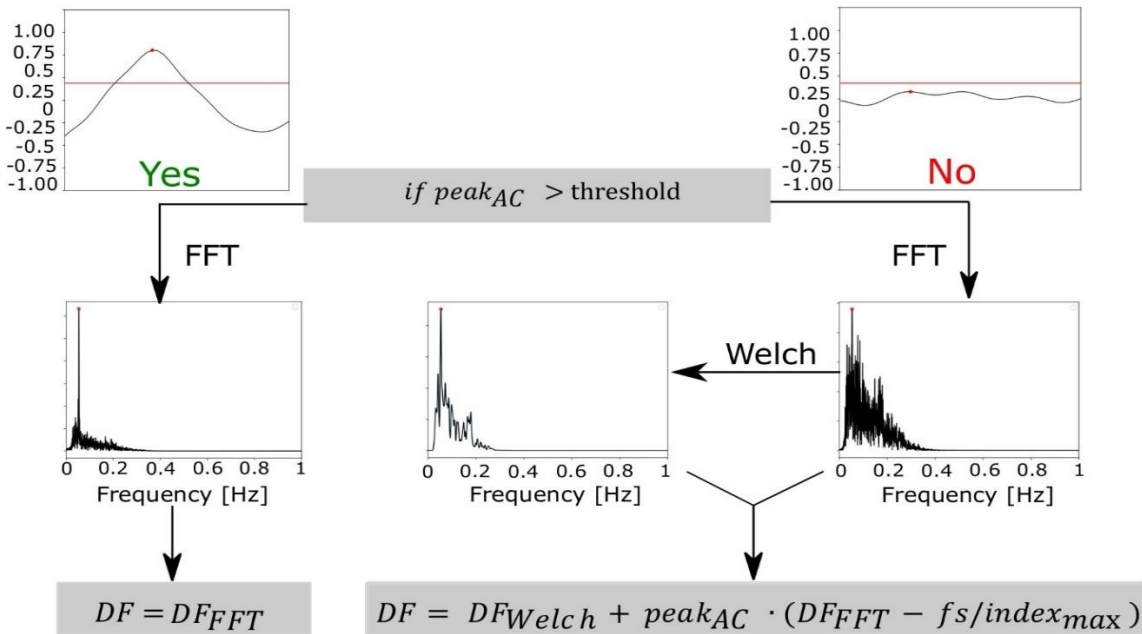

Figure 2. The algorithm illustration for the novel auto-method (NCAM). The threshold is represented by line at value of 0.4 for autocorrelation. If the autocorrelation reaches this threshold ("Yes" on the left panel), then DF is calculated only by application of FFT ($DF_{FFT}$). If not ("No" on the right panel), the DF is calculated by the proposed formula, where $DF_{Welch}$ presents DF calculated using Welch's method, $peak_{AC}$ presents the peak value of normalized autocorrelation, $f_s$ is sampling frequency, and $index_{max}$ is the lag in autocorrelation. In order to illustrate two signals with more and less prominent dominant frequency (DF), we used EGG signals from database [15] from subject ID9 (left panel with $peak_{AC} > 0.4$) and from subject ID6 (right panel with $peak_{AC} \leq 0.4$). Both signals were recorded from CH2 and for postprandial states.

As Welch's method uses averaged periodograms we hypothesized that it will be least vulnerable to artifacts and that it should be dominant method in NCAM in the case of less prominent FFT peak. Then, we also added a scaled difference between dominant frequency calculated by application of FFT and autocorrelation, where scale corresponds to the peak of the normalized autocorrelation (Fig. 2). We hypothesized that the correction of the DF with this scaled difference would yield to more accurate output value. The threshold value of 0.4 was chosen empirically, after a careful visual inspection of the signals and applied trial and error procedure.

## 2.3 Comparison of presented methods and statistical tests

In order to test the algorithms in different conditions, the DF was calculated in relation to the following:

1. fasting and postprandial,
2. three electrode locations (CH1, CH2, and CH3), and
3. low BMI group, high BMI group, and for all subjects.

For upper three testing conditions, we presented average DFs with corresponding SDs in Table and graphically by box plots.

A paired-sample t-test was used in order to verify whether the dominant frequencies in fasting and postprandial cases were significantly different, as described in [14]. Results that had p-value < 0.05 were considered statistically significant.

We calculated the relative difference $R_{def}$ between DF calculated with the proposed method ($DF_{method}$) where method presents Welch's spectrogram, FFT, autocorrelation, and novel combined procedure NCAM, and between $DF_{benchmark}$ i.e., the dominant frequency of the benchmark data by the following equation:

$$R_{def} = \frac{|DF_{benchmark} - DF_{method}|}{DF_{benchmark}} \cdot 100\% \tag{6}$$

In order to test whether NCAM procedure can differentiate between low and high BMI groups, we counted and reported the number of data records with normalized autocorrelation that exceeded threshold of 0.4.

In order to test the robustness of the proposed NCAM procedure and separate methods, we used additive Gaussian noise with the EGG signal recorded in subject ID9 on CH2 during postprandial state. This EGG record was marked as "clean" i.e., artifact-free by visual inspection. Gaussian noise was added to the signal in a SNR range from -40 dB to 20 dB in a step of 2 dB. Overall, 30 semi-synthetic waveforms were formed. For each method and all semi-synthetic EGG signals, we calculated relative difference given in Relation (6) and compared algorithms performance.

# 3 Results

When for each *S*, paired-sample t-test was used to compare the two states (fasting and postprandial), the closest p-values to the ones obtained in Popović et al. [14] were used to select optimal window length and *N*/4 was obtained for CH1 and CH2, and CH3 combined.

Table 1. The number of recordings that displayed peak > 0.4 in normalized auto-correlation when divided into BMI (Body Mass Index) groups and electrode positions (CH1, CH2, and CH3).

|  | High BMI (BMI > 25 kg/$m^2$) | | | Low BMI (BMI < 25 kg/$m^2$) | | |
|---|---|---|---|---|---|---|
|  | CH1 | CH2 | CH3 | CH1 | CH2 | CH3 |
| peak > 0.4 | 1 | 1 | 0 | 7 | 8 | 6 |
| Total | 10 | 10 | 10 | 10 | 10 | 10 |

Table 2. Averaged DFs with standard deviations in cpm (cycles per minute) for two BMI groups (low and high) and for all subjects, for EGG signals recorded from three recording locations (CH1, CH2, and CH3), and for two states (fasting and postprandial). DFs are presented for four methods: AC – autocorrelation, Welch – Welch's method, FFT – Fast Fourier Transform, and NCAM – proposed combined procedure.

| BMI group | Channel | State | DF (AC) | DF (Welch) | DF (FFT) | DF (NCAM) |
|---|---|---|---|---|---|---|
| All | CH1 | Fasting | 2.88 ± 0.29 | 2.71 ± 0.33 | 2.75 ± 0.38 | 2.69 ± 0.36 |
|  |  | Postprandial | 2.99 ± 0.36 | 2.90 ± 0.49 | 3.11 ± 0.49 | 2.93 ± 0.49 |
|  | CH2 | Fasting | 2.93 ± 0.35 | 2.69 ± 0.38 | 2.72 ± 0.39 | 2.65 ± 0.40 |
|  |  | Postprandial | 3.01 ± 0.32 | 3.06 ± 0.49 | 3.00 ± 0.36 | 3.07 ± 0.50 |
|  | CH3 | Fasting | 2.97 ± 0.47 | 2.76 ± 0.37 | 2.80 ± 0.36 | 2.73 ± 0.40 |
|  |  | Postprandial | 2.95 ± 0.39 | 2.94 ± 0.42 | 3.01 ± 0.36 | 2.95 ± 0.42 |
| Low BMI | CH1 | Fasting | 2.78 ± 0.24 | 2.67 ± 0.26 | 2.74 ± 0.32 | 2.66 ± 0.28 |
|  |  | Postprandial | 3.13 ± 0.21 | 2.89 ± 0.40 | 3.07 ± 0.24 | 2.88 ± 0.40 |
|  | CH2 | Fasting | 2.84 ± 0.21 | 2.78 ± 0.31 | 2.75 ± 0.30 | 2.76 ± 0.33 |
|  |  | Postprandial | 3.12 ± 0.19 | 3.18 ± 0.55 | 3.04 ± 0.22 | 3.18 ± 0.55 |
|  | CH3 | Fasting | 2.79 ± 0.20 | 2.78 ± 0.27 | 2.78 ± 0.23 | 2.77 ± 0.28 |
|  |  | Postprandial | 3.02 ± 0.25 | 3.06 ± 0.22 | 3.04 ± 0.22 | 3.07 ± 0.23 |
| High BMI | CH1 | Fasting | 2.99 ± 0.31 | 2.75 ± 0.39 | 2.71 ± 0.42 | 2.72 ± 0.42 |
|  |  | Postprandial | 2.84 ± 0.42 | 2.91 ± 0.57 | 3.09 ± 0.63 | 2.97 ± 0.56 |
|  | CH2 | Fasting | 3.01 ± 0.42 | 2.60 ± 0.43 | 2.62 ± 0.43 | 2.54 ± 0.43 |
|  |  | Postprandial | 2.92 ± 0.38 | 2.95 ± 0.40 | 2.91 ± 0.44 | 2.95 ± 0.41 |
|  | CH3 | Fasting | 3.16 ± 0.59 | 2.74 ± 0.45 | 2.75 ± 0.44 | 2.69 ± 0.48 |
|  |  | Postprandial | 2.88 ± 0.49 | 2.82 ± 0.53 | 2.92 ± 0.43 | 2.83 ± 0.52 |

The number of the recordings in which autocorrelation peak was higher than 0.4 and overall number of tested EGG waveforms in relation to the BMI groups is presented in Table 1.

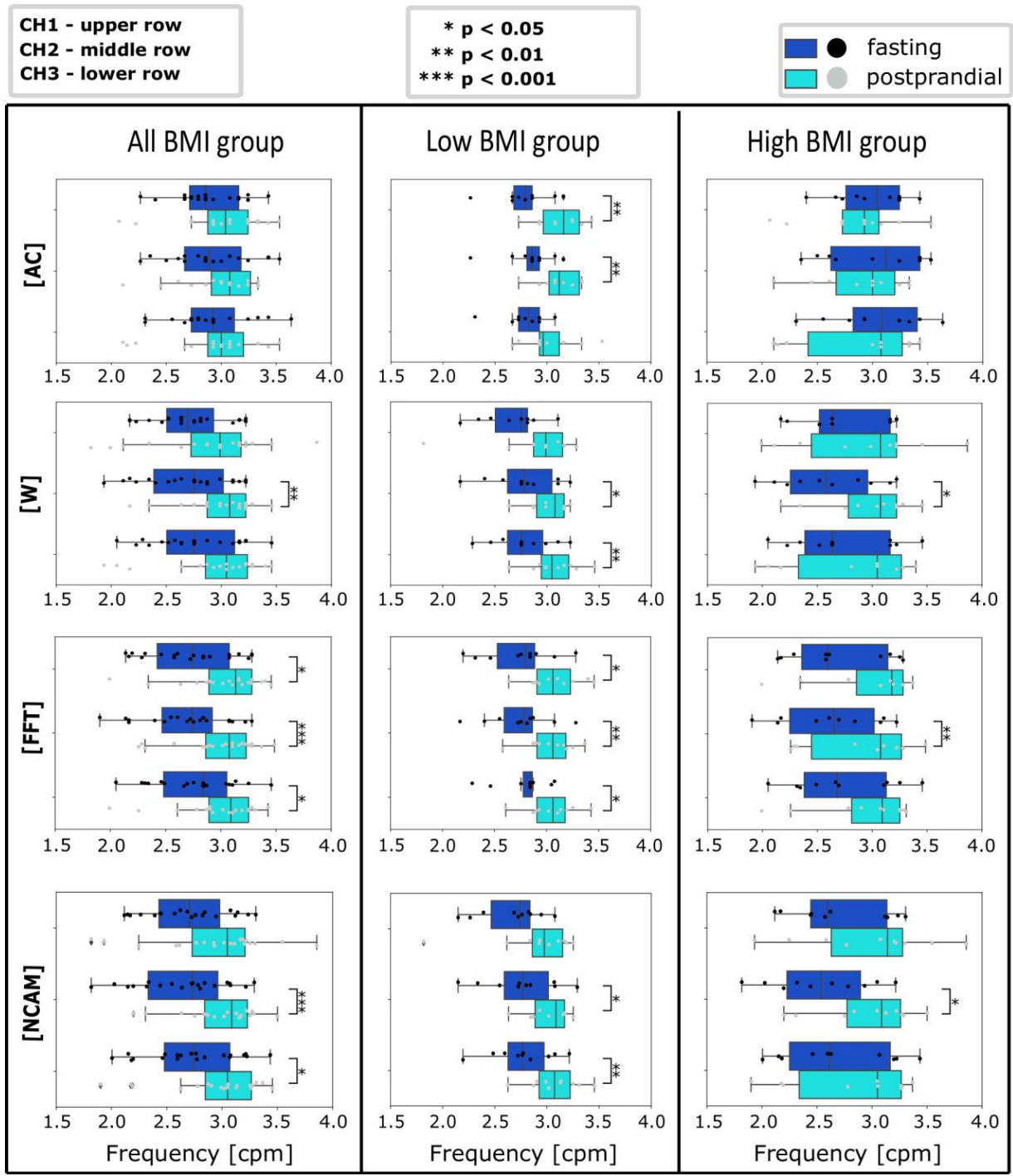

Figure 3. Box plot of dominant frequencies (DF) with individual samples during fasting and postprandial states for four methods for DF calculation, AC - autocorrelation, W - Welch's method, FFT - Fast Fourier Transform, NCAM: our method, and for three groups of subjects (all BMI, low BMI, and high BMI groups).

Average DFs with SDs for all subjects and two groups (low BMI and high BMI), for EGG signals recorded in CH1, CH2, and CH3 locations during fasting and postprandial conditions for methods used to calculate DF are presented in Table 2.

In Fig. 3, box plots for all measurement conditions and all applied methods for DF calculation are presented together with results of statistical test expressed in three categories of p values.

Statistically significant difference between fasting and postprandial states in all three groups of subjects (all, low BMI, and high BMI) was achieved only on CH2 recording location (as shown in benchmark data) and with FFT, Welch's method, and our proposed NCAM approach (Fig. 3).

Relative differences between calculated DF and benchmark DF obtained for different SNRs in semi-synthetic EGG signal by application of following methods for DF calculation: autocorrelation, FFT, Welch's method, and proposed NCAM solution are presented in Fig. 4.

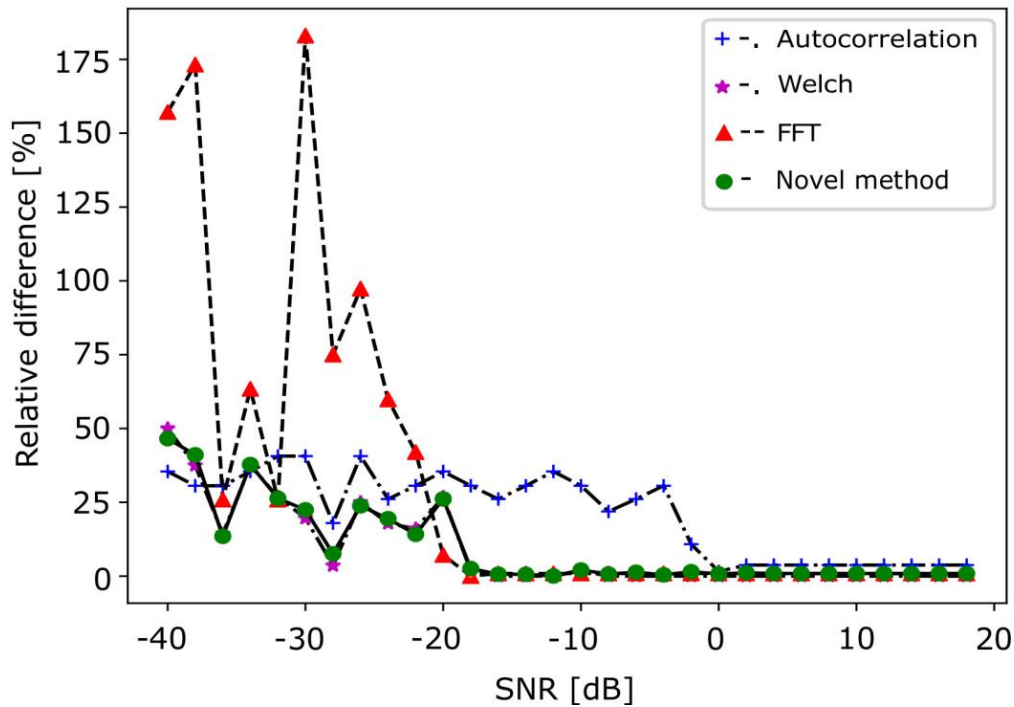

Figure 4. The relative differences for calculated DF using auto-correlation, FFT, Welch's method, and NCAM (novel method) versus benchmarks DF depending on SNR in a range from -40 dB to 20 dB with step of 2 dB when applied on a synthetic dataset derived from EGG and additive white Gaussian noise in subject ID9 for recordings in CH2 and during postprandial state.

# 4 Discussion

Calculation of DF on the EGG data set by the application of different methods confirmed that there could be a critical variation in obtained values (see Fig. 3 and Table 2) for the same recordings. Consequently, selection of the algorithm can highly influence conclusions derived from the EGG recording. It is suggested that visual inspection by educated observer is needed for suitable analysis of EGG [26], which can be time consuming and challenging for the inexperienced researchers and physicians. This inspired as to assess performance of different automated algorithms, and subsequently propose a novel one.

Autocorrelation showed great sensitivity to BMI (Fig. 3 and Tables 1-2). In the high BMI group, mean DF was greater in fasting than in postprandial state on all three channels (2.99 cpm, 3.02 cpm, 3.16 cpm in fasting and 2.85 cpm, 2.90 cpm, 2.88 cpm in the postprandial state for CH1, CH2, CH3, respectively). Numerous studies have shown that the postprandial DF is slightly higher than the fasting one [27-29]. In the study by Tolj et al. [30], subjects in the BMI range 25-30 kg/m² had unexpectedly higher fasting DFs than postprandial ones which is in agreement with the results that we obtained by the application of autocorrelation in Table 2. However, results obtained by the means of autocorrelation should be taken with a grain of salt, as we obtained similar unpredicted result in all subjects by application of autocorrelation. The reason for this erroneous DF calculation might be the spectral resolution of autocorrelation methods which is much lower than in FFT or Welch's method. To the authors' knowledge, the autocorrelation method has not yet been used in the relevant literature for EGG signal processing.

One of the main parameters when applying Welch's spectrum is the window length. It is chosen mainly in an empirical fashion. With the determined window size of N/4, and the recording time of 20 min, optimal window size was comprised of a 5 min interval (300 s). Commonly used tool for EGG power spectrum analysis is the running spectrum [31] where Sanmiguel et al. used 240 s interval for the running spectrum analysis [27], while Pfaffenbach et al. used a 256 s interval [27], which is consistent with our result. Calculated values of DF by Welch's method supported increase in slow wave frequency after meal intake. Statistically significant differences were obtained for CH2 in both low and high BMI groups, as well as for all subjects, which supports supremacy of CH2 recording position compared to CH1 and CH3 which is in agreement with results in [14]. Lower p values than the ones from benchmark data [14] could be the consequence of decreased spectral resolution and absence of manual correction.

Running spectral analysis, as the commonly used method for the calculation of DF, is based on FFT [26]. Results presented in Fig. 3. suggest that on analyzed EGG data, even without manual correction, there is a statistically significant postprandial increase in DF. In the terms of p value, FFT outperformed our novel algorithm. Explanation of this can lie in the fact that the EGG data was recorded for relatively short time frame, in static environment, and in healthy subjects, so it was expected to acquire almost noise-free signals with clear dominant peak. The question arises – Would the result be different if the signals are highly compromised with various artifacts? In order to answer that, additional analysis was performed.

Researchers showed that BMI does not affect the DF, but implied that higher BMI decreases the SNR in the recorded signal [13, 29, 30]. The group in which spectral peaks could not be determined had a significantly higher BMI than the group with a well-defined spectral peak [13]. Our results support this claim, as only 2 (of 30 recordings) in the higher BMI group had normalized autocorrelation peak higher than 0.4 (and therefore a well-defined spectral peak), while 21 in the lower BMI group displayed the same feature (Table 1). Even though BMI does not affect the DF value [29], it affects other characteristics of the EGG signals, such as signal amplitude which is decreased in high BMI group, that has been explained by the greater stomach pacemaker distance from the abdominal wall [30]. We showed that subjects with BMI > 25 kg/m² displayed lesser peak prominence in normalized autocorrelation, which can be attributed to lower SNR compared to BMI < 25 kg/m² subjects. In regard to the obtained results, we proposed a method that combines all aforementioned signal processing methods. It displayed better results over autocorrelation because it was more accurate, especially in the -20 dB to -10 dB SNR band (Fig. 4). The accuracy was measured as the relative difference between DFs obtained using the specific method and benchmark, manually corrected data. It outperformed FFT due to its better noise resistance, and it was better than simply using Welch's method because it retained the global features from the signal and thus giving statistically significant results just as FFT (Fig. 3). Several methods for the automatic assessment of the signal quality were proposed [13, 32]. Wolpert et al. [13] devised an algorithm for automatically discarding the bad EGG recordings in healthy subjects. However, their method assumes that the subject is healthy as one of the discarding conditions is that signal is less than 70% in the normogastric range. They also emphasize the need for visual inspection due to signal distortion after filtering. Espinosa et al. suggested an automatic signal labeling based on EGG amplitude. The signals were visually labeled by an expert and then used as benchmark data. They obtained an average coincidence of 97.1% for patients with diabetes and 88.6% for healthy volunteers [32]. Presented NCAM method shows great promise for future work, as it achieved the same DF values as the FFT of the manually corrected data while having better resistance to noise than the FFT itself (Fig. 4).

The autocorrelation and PSD functions of a wide-sense-stationary random process are connected via the Wiener-Khinchin theorem [33, 34]. We perceived that the higher autocorrelation peak implies a well-defined spectral peak. This could prove useful in future work for the analysis of EGG signal quality, as normalized auto-correlation is cheaper in computing power than the PSD function.

Limitations of the presented study are:

1. The used dataset did not include subjects with BMI > 30 $\frac{kg}{m^2}$.

2. The conclusions presented here should be checked on a larger sample and in different recording conditions, as well as in patients with gastric disorders.

3. We presented noise only with the additive white Gaussian process in the synthetic dataset being a simplified representation of possible EGG artifacts [35].

4. Our novel method is still not efficient enough when it comes to very low SNR (<-30 dB).

5. We did not test the proposed NCAM algorithm for different sampling frequencies and various time and frequency resolutions when calculating spectral parameters which could possibly disturb obtained results.

# 5 Conclusions

In this paper, we presented a novel auto-method for calculating dominant frequency in healthy subjects. We showed that the accuracy in a noisy environment can be drastically improved, by combining three methods: autocorrelation, FFT, and Welch's method for spectral density estimation. The method still captures the global aspects of the signal and retains the statistical significance when applied on data recorded in fasting and postprandial states in comparison with benchmark data.

# Author contributions

**N. Jovanović**: Conceptualization, Investigation, Software, Visualization, Writing- Original draft preparation, **N. B. Popović**: Conceptualization, Methodology, Validation, Writing - review & editing, **N. Miljković**: Conceptualization, Methodology, Supervision, Writing - review & editing.

# Funding sources

N. Miljković was partly supported by the Ministry of Education, Science and Technological Development, Republic of Serbia (MEST), Grant No. TR33020. MEST had no involvement in study design, in collection, in analysis and interpretation of data, in writing of the report, and in the decision to submit the article for publication.

# Declarations of Competing Interest

The authors declare that they have no known competing financial interests or personal relationships that could have appeared to influence the work reported in this paper.